\documentclass[prd,superscriptaddress,twocolumn,showpacs,preprintnumbers,nofootinbib,amsmath,amssymb]{revtex4}

\usepackage{graphicx}
\usepackage{dcolumn}
\usepackage{bm}

\newcommand{\cmsec}{{\mathrm{cm}^{-2}~\mathrm{s}^{-1}}}

\newcommand{\second}{{\mathrm{s}}}
\newcommand{\GeV}{\mathrm{GeV}}
\newcommand{\pcs}{\mathrm{pc}}
\newcommand{\sr}{\mathrm{sr}}
\newcommand{\Mmax}{M_{\mathrm{max}}}
\newcommand{\Mmin}{M_{\mathrm{min}}}

\begin{document}

\title{Can proper motions of dark-matter subhalos be detected?}

\author{Shin'ichiro Ando}
\author{Marc Kamionkowski}
\author{Samuel K.~Lee}
\affiliation{California Institute of Technology, Mail Code 130-33,
Pasadena, CA 91125}

\author{Savvas M. Koushiappas}
\affiliation{Department of Physics, Brown University, 182 Hope St.,
Providence, RI 02912}

\date{September 4, 2008; accepted October 17, 2008}

\begin{abstract}
One of the goals of NASA's Fermi Gamma-ray Space Telescope
(formerly GLAST) will be detection of gamma rays from dark-matter
annihilation in the Galactic halo.  Theoretical
arguments suggest that dark matter may be bound into subhalos
with masses as small as $10^{-4}$--$10^{2}\, M_\oplus$.  If so, it
may be possible to detect individual subhalos as point sources in
the Fermi Telescope.  It has further been argued that some of these point
sources may exhibit proper motions.  Here we show that
upper limits to the diffuse gamma-ray background constrain
the number of subhalos close enough to exhibit proper motions to be
less than one.
\end{abstract}


\maketitle

\section{Introduction}
\label{sec:intro}

One of the major goals of the recently launched Fermi Gamma-ray
Space Telescope (formerly GLAST) \cite{glast} is to detect dark-matter
annihilation into gamma rays, as
expected, with some flux, if dark matter is composed of 
weakly-interacting massive particles (WIMPs), such as the
neutralino in supersymmetric models~\cite{DMreviews}.
Theoretical arguments suggest that dark matter in the Galaxy may
be distributed in subhalos~\cite{substructure, substructure2}
with masses that may extend all the way down to 10$^{-4}$--10$^2 \,
M_{\oplus}$, if WIMPs make up dark matter~\cite{Profumo:2006bv}.

References~\cite{propermotion} pointed out that if WIMP substructure
extends to such small scales, then it may be possible not only to detect
individual subhalos as point sources with the Fermi Telescope, but also
that up to ten of these point sources may exhibit proper motions.

In this paper, we revisit this proposal.
We use the diffuse gamma-ray background measured with the Energetic
Gamma Ray Experimental Telescope (EGRET)~\cite{egret}
to place an upper limit to the gamma-ray luminosity of these subhalos
(see also Ref.~\cite{OTM}).
This constraint then limits the number of detectable subhalos
that are sufficiently close to exhibit proper motions.
A similar argument was also made in Ref.~\cite{Pieri} for a
specific numerically-simulated subhalo model.
Here, we present an analytic and fully general argument.
The largest number of detectable proper motions is obtained if
the subhalo mass distribution is sharply peaked around
$10^{-3}$--$10^{-2}\, M_{\odot}$, and even then, the expected number of
detectable proper motions is less than 1.

Before proceeding with the detailed calculation, we begin in
Sec.~\ref{sec:OoM} with an order-of-magnitude estimate.
Section~\ref{sec:egret} then describes more precisely the EGRET
upper limit to the diffuse background and the constraints
implied for the subhalo population.  We then evaluate in
Sec.~\ref{sec:proper motion} the number of detectable proper
motions, first by assuming all subhalos have the same mass,
and then by generalizing to the more
realistic case of a power-law distribution of subhalo masses.
In Sec.~\ref{sec:conclusions}, we discuss the results and how
they vary upon changing the assumptions taken from their 
canonical values adopted in the main body of the paper. 

\section{Order of Magnitude Estimate}
\label{sec:OoM}

If the Fermi Telescope can detect proper
motions as small as $0.1^\circ$ and subhalos have a transverse
velocity $\sim$200~km~s$^{-1}$, then a subhalo must be within
a distance $d_{\mathrm{pm}}\sim 0.35$~pc to have a detectable
proper motion over three years.  If the point-source flux sensitivity is
$F_{\mathrm{sens}}$, a source with a detectable proper motion
must have a luminosity $L\sim 4\pi d_{\mathrm{pm}}^2
F_{\mathrm{sens}}$.  Each detected proper motion thus implies a
subhalo contribution to the luminosity density of order
$L/V \sim 3 F_{\mathrm{sens}}/d_{\mathrm{pm}}$.  This implies a
total luminosity, integrated
over the volume of the Milky Way interior to the Solar radius,
of $L_{\mathrm{MW}} \gtrsim 4 \pi r_\odot^3 F_{\mathrm{sens}}
/d_{\mathrm{pm}}$, where $r_\odot=8.5$ kpc is our distance from the
Galactic center (the ``$\gtrsim$'' arises because the halo density
most generally decreases with radius).  Thus, a flux from the
Milky Way of $F_{\mathrm{MW}} \sim L_{\mathrm{MW}}/(4\pi r_\odot^2)
\gtrsim (r_\odot/d_{\mathrm{pm}}) F_{\mathrm{sens}}$ is implied for
each proper motion detected.  This is a {\it large} flux---larger
than upper limits even for one detected proper motion---as the
more detailed calculations below will show.

\section{EGRET constraints to the subhalo population}
\label{sec:egret}

Assume that a fraction $f$ of the dark matter in the Galactic halo
is composed of subhalos of mass $M$ and luminosity
$L$,\footnote{Throughout this paper, the luminosity is the {\it
number} (not energy) of photons emitted per unit time;
similarly, we deal with number fluxes and intensities.} and we
define a mass--to--gamma-ray-luminosity ratio $\Upsilon\equiv
M/L$.  Assume further that the radial distribution of subhalos
follows the density profile $\rho(r)$ of the Milky Way halo.

The gamma-ray intensity $I(\psi)$ (units of
photons~$\cmsec~\sr^{-1}$), due to gamma rays from these subhalos,
is
\begin{equation}
      I(\psi) =  \frac{f}{4 \pi \Upsilon} \int
        dl(\psi)\ \rho\left(r[l(\psi)] \right),
\label{eqn:intensity}
\end{equation}
along a line of sight at an angle $\psi$ from the Galactic
center.  Here, $l$ is the distance along the line of sight
$\psi$; i.e., $r^2=r_\odot^2+l^2 -2 r_\odot l \cos\psi$.

For $\rho(r)$, we use the Navarro-Frenk-White (NFW)~\cite{NFW96}
profile,
\begin{equation}
     \rho(r) = \frac{\rho_s}{(r/r_s) (1+r/r_s)^2},
\label{eqn:NFW}
\end{equation}
where $\rho_s = 5.4 \times 10^{-3} M_{\odot}~\mathrm{pc}^{-3}$
is the characteristic density, and $r_s = 21.7$ kpc is the scale
radius.
(These numerical values, taken to explain the virial mass and
rotation curves~\cite{Fornengo}, are adopted for illustrative
purposes, but the main conclusions do not depend on them.)
This gives $\rho_\odot = 7\times 10^{-3}
M_\odot$ pc$^{-3}$ as the local density.

The diffuse gamma-ray background measured with EGRET places an
upper limit to $f\Upsilon^{-1}$.
If the subhalo gamma rays form a continuum spectrum, then we 
use the diffuse gamma-ray-background intensity measured with
EGRET~\cite{EGRET} at $\psi=90^\circ$ as a conservative upper
limit to $I(\psi)$ from subhalos (roughly the same bound obtains
for $\psi=180^\circ$).
For line emission, on the other hand, we approximate the upper limit to
the gamma-ray-line intensity, averaged over a $10^\circ\times10^\circ$
region around the Galactic center, by $2 \times
10^{-6}\,(E/\GeV)^{-1/2}\, \cmsec~\sr^{-1}$ over the energy range
$0.1\,\GeV < E <10\, \GeV$~\cite{Pullen:2006sy}.
(Scaling our results to more conservative limits on line
intensity~\cite{Mack} is trivial.)
At an energy $E = 10\, \GeV$, the upper
limits to both the continuum and line intensities turn out,
coincidentally, to be
\begin{equation}
  f\Upsilon^{-1} = 10^{29}\,   I_\ast \, M_\odot^{-1}\, \second^{-1},
\label{eqn:LMupperlimit}
\end{equation}
where $I_\ast \leq 1$ is the fractional contribution from
subhalos to the gamma-ray-background intensity, the total being
due, additionally, to traditional astrophysical sources and/or
WIMP annihilation in the $1-f$ fraction of the Milky Way
halo that is smoothly distributed. We discuss further this assumption 
later on.

\section{Proper-motion detection}
\label{sec:proper motion}

Suppose that a dark-matter subhalo moves transverse to the line of
sight with a velocity 200~km~s$^{-1}$.
Then, in a three-year experiment, the subhalo will move an angular
distance of $0.1^\circ\,\theta_{0.1}$ if
it is at a distance
\begin{equation}
   d_{\rm pm} \equiv 0.35\,\pcs\, \theta_{0.1}^{-1}.
\end{equation}
Here, we take the minimum proper motion detectable by the Fermi
Telescope to be $0.1^\circ$ (for 10 GeV gamma rays) as a canonical
value~\cite{LAT}, but keep the $\theta_{0.1}$ dependence in the
following.
Thus, a proper motion can be detected only if the subhalo is
closer than $d_{\rm pm}$.

\subsection{The mono-luminous case}

First, consider the case where the background is due to subhalos of 
a one mass scale $M$. We call this the mono-luminous case. 
The number of subhalos within a distance $r$ of the Earth is
\begin{eqnarray}
 N_{\rm sh} (< r) &=& \frac{4 \pi r^3}{3}
 \frac{f\rho_\odot}{M}
 \nonumber\\
 &=& 1.3 \times 10^{-3} \theta_{0.1}^{-3}
 \left(\frac{r}{d_{\rm pm}}\right)^3
 \left(\frac{M}{f M_\odot}\right)^{-1}.
 \label{eq:subhalo number}
\end{eqnarray}
For the source to be detected at $d_{\rm pm}$, the 
luminosity $L = \Upsilon^{-1} M$ has to be larger than $4 \pi d_{\rm
pm}^2 F_{\mathrm{sens}}$, where $F_{\mathrm{sens}}$ is the point-source flux
sensitivity; for the Fermi Telescope, we take
$F_{\mathrm{sens}}\equiv$10$^{-10}\,F_{10}\, \cmsec$ at 10 GeV,
with $F_{10}\simeq 1$~\cite{LAT}.  With the EGRET constraint to $\Upsilon$
[Eq.~(\ref{eqn:LMupperlimit})], the condition for detection at
$d_{\rm pm}$ is $M \ge M_{\rm pm}$, where
\begin{equation}
  M_{\rm pm} \equiv 1.5 \times 10^{-2} \, f M_\odot F_{10}
  I_\ast^{-1} \theta_{0.1}^{-2}.
  \label{eq:Mpm}
\end{equation}
If $M < M_{\rm pm}$, the subhalo can only be
detected up to a distance $[\Upsilon^{-1} M / (4 \pi F_{\rm
sens})]^{1/2}<d_{\rm pm}$.

Thus, the number of detectable proper motions is
\begin{eqnarray}
  N_{\rm pm} &=& \left\{\begin{array}{ccc}
  N_{\rm sh}(<d_{\rm pm}), & \mbox{for} & M \ge M_{\rm pm}, \\
  N_{\rm sh}\left(< \left[\frac{\Upsilon^{-1} M}{4 \pi F_{\rm sens}}
  \right]^{1/2} \right), & \mbox{for} & M < M_{\rm pm},
  \end{array}\right.
  \nonumber\\
  &=& 0.09 F_{10}^{-1} I_\ast \theta_{0.1}^{-1}
  \times \left\{\begin{array}{ccc}
  \left(\frac{M}{M_{\rm pm}}\right)^{-1}, & \mbox{for} & M \ge M_{\rm pm},
  \\
  \left(\frac{M}{M_{\rm pm}}\right)^{1/2}, & \mbox{for} & M < M_{\rm pm}.
  \end{array}\right.
  \nonumber\\
  \label{eq:Npm mono-luminous}
\end{eqnarray}
To summarize, assuming the diffuse background density measured by EGRET, and 
assuming that this background is due to annihilations in subhalos, 
the maximum probability to detect proper motion in a nearby subhalo  
is $\sim$10\%. This probability is maximized
for subhalo masses $M \sim 10^{-2}\, f\, M_\odot$.
Larger-mass subhalos will provide a larger number of detectable
point sources (that can be detected to larger distances), but
fewer with proper motions. On the other hand, there will be more 
nearby objects if
they are less massive, but they will be too faint to be detectable.

\subsection{Power-law mass and luminosity functions}
\label{sec:mass function}

Now suppose, as before, that a fraction $f$ of the mass of the
halo is in subhalos, but now assume that the masses are
distributed with a power-law  mass function,
\begin{eqnarray}
  \frac{dn}{dM}(r,M) &=& \frac{f \rho(r)}{\ln \Lambda} M^{-2},
  \label{eq:mass function}
\end{eqnarray}
with power-law index $-2$, as suggested by numerical
simulations~\cite{substructure2}.  Here, $\Lambda \equiv \Mmax /
\Mmin$, and $\Mmin$ and $\Mmax$ are the lower and upper mass
limits.  If the population has a constant mass-to-luminosity
ratio $\Upsilon$, then the bound in Eq.~(\ref{eqn:LMupperlimit})
still applies.

The differential number
$dN_{\rm sh} / d\ln M$ of subhalos per logarithmic mass
interval within a distance $r$ from the Earth is given by
Eq.~(\ref{eq:subhalo number}), with the replacement $f \to f /
\ln \Lambda$.  The simulations of
Refs.~\cite{substructure2} suggest $f / \ln \Lambda = 0.0064$;
typical values are $\ln \Lambda = 35$ and $f \approx 0.2$.
The differential number of subhalos with detectable flux and
proper motions per logarithmic mass interval is then
\begin{eqnarray}
  \frac{dN_{\rm pm}}{d\ln M} &=&
  0.002 F_{10}^{-1} I_\ast \theta_{0.1}^{-1} \left(\frac{\ln
  \Lambda}{35}\right)^{-1}
  \nonumber\\&&{}\times
  \left\{\begin{array}{ccc}
  \left(\frac{M}{M_{\rm pm}}\right)^{-1}, & \mbox{for} & M \ge M_{\rm pm},
  \\
  \left(\frac{M}{M_{\rm pm}}\right)^{1/2}, & \mbox{for} & M < M_{\rm pm}.
  \end{array}\right.
  \label{eq:dNdM}
\end{eqnarray}
Again the largest contribution comes from subhalos with masses around
$M_{\rm pm}$.
By integrating this expression over $\ln M$, we have
\begin{eqnarray}
  N_{\rm pm} & \approx &
  0.002 F_{10}^{-1} I_\ast \theta_{0.1}^{-1} \left(\frac{\ln
  \Lambda}{35}\right)^{-1}
  \nonumber\\&&{}\times
  \left\{\begin{array}{ccc}
  \left(\frac{\Mmin}{M_{\rm pm}}\right)^{-1}, & \mbox{for} &
  M_{\rm pm} \le \Mmin, \\
  3, & \mbox{for} & \Mmin \ll M_{\rm pm} \ll \Mmax,
  \end{array}
  \right.
  \nonumber\\
\label{eqn:logNpm}
\end{eqnarray}
which is far smaller than in the mono-luminous case.  This is
easily understood: Detectable proper motions still come from
subhalos with masses around $M_{\rm pm}$.  However, there are
now fewer such subhalos, since the  subhalo masses are now
distributed over a wider range.  The contribution from
$M_{\rm pm}$ is thus smaller by a logarithmic factor $\ln \Lambda$.
Although we focused only on the case of $dn / dM \propto M^{-2}$ here,
the argument applies to any mass function.  Again, the largest
number of detectable proper motions comes when the mass
distribution is sharply peaked around $M_{\rm pm}$.

\section{Discussion and conclusions}
\label{sec:conclusions}
 
In this short paper, we revisit the detectability of proper motions of
dark-matter subhalos. Our main results are: (1) The EGRET diffuse-background
constraint on the subhalo mass-to-luminosity ratio severely
restricts the expected number of proper-motion detections with
the Fermi Telescope to be less than 1.  This argument does not rely on
the details of any particle-physics model, in contrast to the
earlier studies~\cite{propermotion}.  (2) The number
of detectable proper motions is maximized if the subhalo mass 
distribution is sharply peaked around a mass $M_{\rm pm}\sim
0.01\, f M_\odot$.  In this case, the probability to
detect one subhalo with a proper motion is $\sim$10\%.  
(3) This maximal probability is obtained only when
dark-matter annihilation from subhalos dominates the diffuse gamma-ray
background (i.e., $I_\ast\simeq 1$), and it is reduced in
proportion to the fraction $I_\ast$ of the diffuse background
due to subhalos.  This fraction is likely to be much less than
1, as there may be significant contributions to EGRET's diffuse
background from unresolved astrophysical sources, and perhaps
from WIMP annihilation in a smoothly distributed halo component.
(4) If the subhalo mass
function follows a power-law mass function, the expected number
of proper-motion detections is further suppressed by a
logarithmic factor $\ln (\Mmax / \Mmin)$, where $\Mmax$ and
$\Mmin$ are masses of the largest and smallest subhalos.  This
factor could be $\sim$30--40, in which case the chance of
detection would be no greater than 1\%.
In this calculation, we have used a canonical gamma-ray energy of 
$E=10$~GeV.  The results can be scaled (but will not differ too
much) by taking into account the $E^{-1/2}$ dependence of
$I_\ast$ and a roughly similar energy dependence of the Fermi Telescope
point-source sensitivity $F_{10}$.  

Under these assumptions, the main result is that the probability of detecting 
proper motion in subhalos is less than one. It is natural then to ask under 
what conditions proper motions {\it can} be detected. There are two 
parameters in our formalism which can be varied in order to explore this 
possibility. 

The quantity $I_\ast$ can be used to parameterize
the effects of halo-modeling uncertainties. If $\rho(r)$
increases less rapidly toward the Galactic center than in the
NFW profile, then the predicted intensity will decrease and the
bound in Eq.~(\ref{eqn:LMupperlimit}) is weakened (for line
radiation, which is bound by the Galactic center) accordingly.
If, for example, we postulate an extreme scenario where the halo
density $\rho(r)$ is constant interior to the solar radius
(holding the local density fixed), then the bound in
Eq.~(\ref{eqn:LMupperlimit}) is weakened by about a factor of 4.
Likewise, $I_\ast$ can also quantify the effect of a
Galactocentric-radius--dependent fraction $f$.  It may be that
subhalos are more effectively disrupted closer to the Galactic
center, and if so, then we should replace the fraction $f$ in
Eq.~(\ref{eqn:intensity}) by a function $f(r)$ that decreases
monotonically as $r \rightarrow 0$, and move $f(r)$ into the
integrand.  If we guess $f(r)= f_0 (r/r_\odot)^\alpha$, we then
find that for $\alpha = 1$, the bound in Eq.~(\ref{eqn:LMupperlimit})
(with the replacement $f \to f_0$) is weakened by about a factor
of 4 (the bound is weakened further by a factor of 6 if $\alpha =
2$).
Thus, if $f$ has a radial dependence, then its effects can be taken
into account  by allowing an increase in $I_\ast$ by a factor of 4
(for $\alpha=1$) to 6 (for $\alpha=2$).
Note that this caveat applies only for
line radiation (which we have bound by an upper limit to the
gamma-ray flux from the Galactic center), but {\it not} to a
continuum, for which we have used diffuse-background upper
limits away from the Galactic center.

The results are sensitive to the angular resolution and 
duration of the experiment. Throughout this 
calculation we assumed a threshold for proper-motion detection of $0.1^{\circ}$ 
and a timescale of three years.  The number of detectable 
proper motions increases as $\theta_{0.1}^{-1}$ [see
Eqs.~(\ref{eq:Npm mono-luminous}) and (\ref{eqn:logNpm})].
Lowering the threshold for proper-motion detection in a fixed timescale 
can be possible if 
a nearby subhalo is present, and detected at a high signal-to-noise ratio. 
In such a case the angular resolution of the instrument will improve as 
$\sim N_\gamma^{-1/2}$, where $N_\gamma$ is the number of photons
detected from the source. In addition, increasing the timescale of the 
experiment from three to, e.g., 10 years (projected lifetime of the
Fermi Telescope) will increase the probability of detection, as 
proper motion is linear with time. 

In summary, unless the distribution of subhalos within the solar radius is 
severely suppressed, and the Fermi Telescope operates for at least 10 years, 
the expected number of proper-motion detections is less than one.

\begin{acknowledgments}
We thank John Beacom for helpful comments.
Work at Caltech was supported by the Sherman Fairchild Foundation
(SA), DoE DE-FG03-92-ER40701, and the Gordon and Betty Moore
Foundation. SMK acknowledges support from Brown University. 
\end{acknowledgments}


\begin{thebibliography}{}

\bibitem{glast}
  {\tt http://fermi.gsfc.nasa.gov}

\bibitem{DMreviews}
  G.~Jungman, M.~Kamionkowski and K.~Griest,
  Phys.\ Rept.\  {\bf 267}, 195 (1996)
  [arXiv:hep-ph/9506380];
  G.~Bertone, D.~Hooper and J.~Silk,
  Phys.\ Rept.\  {\bf 405}, 279 (2005)
  [arXiv:hep-ph/0404175];
  D.~Hooper and S.~Profumo,
  Phys.\ Rept.\  {\bf 453}, 29 (2007)
  [arXiv:hep-ph/0701197].

\bibitem{substructure}
  S.~Ghigna, B.~Moore, F.~Governato, G.~Lake, T.~Quinn and J.~Stadel,
  Mon.\ Not.\ Roy.\ Astron.\ Soc.\  {\bf 300}, 146 (1998)
  [arXiv:astro-ph/9801192];
  A.~A.~Klypin, A.~V.~Kravtsov, O.~Valenzuela and F.~Prada,
  Astrophys.\ J.\  {\bf 522}, 82 (1999)
  [arXiv:astro-ph/9901240];
  A.~A.~Klypin, S.~Gottl\"{o}ber,A.~V.~Kravtsov and A.~M.~Khokhlov,
  Astrophys.\ J. \ {\bf 516}, 530 (1999);
  B.~Moore, S.~Ghigna, F.~Governato, G.~Lake, T.~Quinn, J.~Stadel and
P.~Tozzi,
  Astrophys.\ J.\  {\bf 524}, L19 (1999);
  A.~Helmi, S.~D.~M.~White and V.~Springel,
  Phys.\ Rev.\  D {\bf 66}, 063502 (2002)
  [arXiv:astro-ph/0201289];
  D.~Reed, F.~Governato, T.~Quinn, J.~Gardner, J.~Stadel and G.~Lake,
  Mon.\ Not.\ Roy.\ Astron.\ Soc.\  {\bf 359}, 1537 (2005)
  [arXiv:astro-ph/0406034];
  A.~Loeb and M.~Zaldarriaga,
  Phys.\ Rev.\  D {\bf 71}, 103520 (2005)
  [arXiv:astro-ph/0504112];
  A.~M.~Green, S.~Hofmann and D.~J.~Schwarz,
  Mon.\ Not.\ Roy.\ Astron.\ Soc.\  {\bf 353}, L23 (2004)
  [arXiv:astro-ph/0309621];
  JCAP {\bf 0508}, 003 (2005)
  [arXiv:astro-ph/0503387];
  J.~Diemand, B.~Moore, and J.~Stadel,
  Nature {\bf 433}, 389 (2005) [arXiv:astro-ph/0501589];
  J.~Diemand, M.~Kuhlen and P.~Madau,
  Astrophys.\ J.\  {\bf 649}, 1 (2006)
  [arXiv:astro-ph/0603250];
  C.~Giocoli, L.~Pieri and G.~Tormen,
  arXiv:0712.1476 [astro-ph];
  V.~Berezinsky, V.~Dokuchaev and Y.~Eroshenko,
  arXiv:0712.3499 [astro-ph];
  M.~Kamionkowski and S.~M.~Koushiappas,
  Phys.\ Rev.\  D {\bf 77}, 103509 (2008)
  [arXiv:0801.3269 [astro-ph]].

\bibitem{substructure2}
  J.~Diemand, M.~Kuhlen and P.~Madau,
  Astrophys.\ J.\  {\bf 657}, 262 (2007)
  [arXiv:astro-ph/0611370];
  Astrophys.\ J.\
  {\bf 667}, 859 (2007)
  [arXiv:astro-ph/0703337].

\bibitem{Profumo:2006bv}
  X.~l.~Chen, M.~Kamionkowski and X.~m.~Zhang,
  Phys.\ Rev.\  D {\bf 64}, 021302 (2001)
  [arXiv:astro-ph/0103452];
  S.~Profumo, K.~Sigurdson and M.~Kamionkowski,
  Phys.\ Rev.\ Lett.\  {\bf 97}, 031301 (2006)
  [arXiv:astro-ph/0603373];
  T.~Bringmann and S.~Hofmann,
  JCAP {\bf 0407}, 016 (2007)
  [arXiv:hep-ph/0612238].

\bibitem{propermotion}
  S.~M.~Koushiappas,
  Phys.\ Rev.\ Lett.\  {\bf 97}, 191301 (2006)
  [arXiv:astro-ph/0606208];
  S.~M.~Koushiappas,
  AIP Conf.\ Proc.\  {\bf 921}, 142 (2007)
  [arXiv:astro-ph/0703778];
  B.~Moore, J.~Diemand, J.~Stadel and T.~R.~Quinn,
  arXiv:astro-ph/0502213.

\bibitem{egret} {\tt http://heasarc.gsfc.nasa.gov/docs/cgro/egret}

\bibitem{OTM}
  T.~Oda, T.~Totani and M.~Nagashima,
  Astrophys.\ J.\  {\bf 633}, L65 (2005)
  [arXiv:astro-ph/0504096].

\bibitem{Pieri}
  L.~Pieri, G.~Bertone and E.~Branchini,
  Mon.\ Not.\ Roy.\ Astron.\ Soc.\  {\bf 384}, 1627 (2008)
  [arXiv:0706.2101 [astro-ph]].

\bibitem{NFW96}
  J.~F.~Navarro, C.~S.~Frenk and S.~D.~M.~White,
  Astrophys.\ J.\  {\bf 490}, 493 (1997)
  [arXiv:astro-ph/9611107].

\bibitem{Fornengo}
  N.~Fornengo, L.~Pieri and S.~Scopel,
  Phys.\ Rev.\  D {\bf 70}, 103529 (2004)
  [arXiv:hep-ph/0407342].

\bibitem{EGRET}
  P.~Sreekumar {\it et al.}  [EGRET Collaboration],
  Astrophys.\ J.\  {\bf 494}, 523 (1998)
  [arXiv:astro-ph/9709257].

\bibitem{Pullen:2006sy}
  A.~R.~Pullen, R.~R.~Chary and M.~Kamionkowski,
  Phys.\ Rev.\  D {\bf 76}, 063006 (2007)
  [arXiv:astro-ph/0610295].

\bibitem{Mack}
  G.~D.~Mack, T.~D.~Jacques, J.~F.~Beacom, N.~F.~Bell and H.~Yuksel,
  Phys.\ Rev.\  D {\bf 78}, 063542 (2008)
  [arXiv:0803.0157 [astro-ph]].

\bibitem{LAT}
  {\tt http://www-glast.slac.stanford.edu}

\end{thebibliography}
\end{document}